\documentclass{singlecol-new}

\usepackage[latin1]{inputenc}
\usepackage{amsfonts}
\usepackage{amsmath}

		\pdfoutput=1 	

	\usepackage[pdftex]{graphicx}

\usepackage{natbib}

\usepackage{xcolor}
\newcommand{\flm}[1]{\textcolor{black}{#1}}

\makeatletter
\newdimen\mathindent
\def\equation{\@beginparpenalty\predisplaypenalty
  \@endparpenalty\postdisplaypenalty
  \refstepcounter{equation}\trivlist \item[]\leavevmode
  \hbox to\linewidth\bgroup $\@eqnnum\m@th\displaystyle\hfil}

\def\endequation{$\hfil\displaywidth\linewidth\egroup\hskip-12pt\endtrivlist}

\makeatother

\newtheorem{exple}{\sc Example}

\newtheorem{defi}{\sc Definition}
\newcounter{defcmpt}
\renewenvironment{defi}[1]{
    \begin{center}
    \begin{minipage}{15cm}
{\flushleft\sc Definition~\thedefcmpt} --- #1

\vspace{0.2cm}
\begin{tabular}{|l}
\begin{minipage}{13cm}
}
{
\end{minipage}\end{tabular}
\stepcounter{defcmpt}
\end{minipage}
\end{center}
}

\newcommand{\N}{\mathbb{N}}
\newcommand{\R}{\mathbb{R}}

\begin{document}

\setcounter{page}{1}

\LRH{Semantic Service Substitution in Pervasive Environments}

\RRH{}

\VOL{?}

\ISSUE{?}

\PUBYEAR{?}

\setcounter{page}{1}

\LRH{Semantic Service Substitution in Pervasive Environments}

\RRH{}

\VOL{?}

\ISSUE{?}

\PUBYEAR{?}

\BottomCatch

\title{Semantic Service Substitution in Pervasive Environments}

\authorA{N. Ibrahim*, F. Le Mouël** and S. Frénot**}

\affA{\flm{*University of Grenoble \\ LIG - Grenoble Informatics Laboratory \\ F-38402 St Martin d'Heres, Grenoble, France \\ **University of Lyon \\ INSA-Lyon, INRIA CITI Lab \\ F-69621 Villeurbanne, France \\ E-mail:
noha.ibrahim@imag.fr, frederic.le-mouel@insa-lyon.fr (corresponding author) and stephane.frenot@insa-lyon.fr
}
}



\begin{abstract}
A computing infrastructure where ``everything is a service''  offers many new system and application possibilities. Among the main challenges, however, is the issue of service substitution for the application execution in such heterogeneous environments. An application would like to continue to execute even when a service disappears, or it would like to benefit from the environment by using better services with better QoS when possible. In this article, we define a generic service model and describe the equivalence relations between services considering the functionalities they propose and their non-functional QoS properties. We define semantic equivalence relations between services and equivalence degree between non-functional QoS properties. Using these relations we propose semantic substitution mechanisms upon the appearance and disappearance of services that fit the application needs in a pervasive environment. We developed a prototype as a proof of concept and evaluated its efficiency over a real use case.
\end{abstract}

\KEYWORD{service-oriented architecture, service substitution, semantic matching, semantic distance, quality of service, z-score, equivalence relations}

\REF{}

\BIO{\flm{\emph{Noha Ibrahim} holds a 'Diplôme d'Ingénieur' from the Ecole Nationale Supérieure d'Informatique et de Mathématique Appliquée de Grenoble (ENSIMAG), and a PhD from National Institute for Applied Sciences (INSA Lyon). Her dissertation focused on providing a spontaneous service integration middleware adapted for pervasive environnements. Noha Ibrahim is currently Associate Professor in the Grenoble Informatics Laboratory (LIG) where she works on service composition framework for optimizing queries and data mining for multimedia applications. \\ \emph{Frédéric Le Mouël} is currently Associate Professor in the National Institute for Applied Sciences of Lyon (INSA Lyon), Telecommunications Department - high-ranked school in France, part of the University of Lyon. He conducts his research in the Center for Innovation in Telecommunication and Integration of Services (INRIA CITI Lab.) where he is leading the Dynamic Software and Distributed Systems for the Internet of Things research group (DynaMid Team). He joined Shanghai Jiao Tong University (SJTU) as Visiting Professor in 2013. His main interests are distributed systems, operating systems, middleware, virtual machines, programming languages, especially in dynamic and autonomic environments. \\ \emph{Stéphane Frénot} holds a 'Diplôme d'Ingénieur' from INSA Lyon, and a PhD from University Lyon I about distributed information systems in hospitals. Stéphane Frénot is currently Professor at the Center for Innovation in Telecommunication and Integration of Services (INRIA CITI Lab.), Telecommunications Department of the National Institute for Applied Sciences of Lyon (INSA Lyon) and in the Rhône-Alpes Complex Systems Institute (IXXI). He is co-heading the INRIA Dice team and is particularly interested in data, web, programming and geopolitics.}}
\maketitle

\section{Introduction} \label{introduction}

A computing infrastructure  (\cite{2})  where ``everything is a service'' offers many new system and application possibilities. Among the main challenges, however, is the issue of service substitution  for the application execution in such heterogeneous environments. An application would like to continue to execute even when a service disappears, or it would like to benefit from the services in the environment by using better services with better quality of service when possible. 

A service publishes a functional interface, describing all the operations that the service can execute. This description is based on semantics and ontologies  (\cite{1}) as pervasive environments (\cite{5,4}) are populated with services from different providers and technologies. Besides the semantic interface description, the interface operations have non-functional properties corresponding to their quality of service. Many middleware and architectures proposed solutions for service substitution  (\cite{7}) or service adaptation  (\cite{9}), but very few described by models, definitions and metrics semantic service substitution adapted for pervasive environments and based on functional interface matching and quality of service computing.
\\
\\
The major contributions of the article are in defining and formalising:

\begin{itemize}
\item The equivalence relations between services considering the functionalities they propose via their functional interfaces. We define and formalise the service model and the service equivalence relations based on the semantic description of their interfaces and operations. Theses relations allow to define if two services are functionally equivalent or not.
\item The QoS degree equivalence functions between the operations and the services. Services can be functionally equivalent but offer and/or require different parameters for quality of service. The QoS equivalence degree gives a metric that indicates how close two services are in terms of their quality of services.
\item Service substitution mechanisms for applications executing in pervasive environments. Based on service equivalence relations and QoS equivalence degree, the pervasive environment can decide to substitute services by functionally equivalent ones, with better QoS computed via the QoS equivalence degree. These service substitution are done transparently and spontaneously as services appear and disappear in the environment with the users coming and leaving.
\end{itemize}

To define, formalise and explain our relations and metrics we adopted the following model. The ``\small{DEFINITION}\normalsize'' paragraphs define the relations and functions between concepts, operations and services using simple grammar and language, whereas the ``\small{EXAMPLE}\normalsize'' paragraphs illustrate and explain these definitions via a use case. 

We begin in section 2 by exposing the state of the art. We define in section 3 the service equivalence relations and the non-functional QoS degree equivalence metrics. We then explain in section 4 the semantic service substitution in pervasive environments. Section 5 details the developed proof-of-concept prototype and its results. Finally, section 6 concludes and gives perspectives for this work.

\section{State of the Art}

In pervasive environments, service substitution (\cite{7,11}) and service similarity problems (\cite{12, 8,13}) have become the new trends in the service-oriented community after the service discovery and service composition problems. Once services are deployed, accessed, executed and composed, the pervasiveness of the environment imposes researchers to find solutions for the service unavailability problem. Indeed, in a pervasive environment, services can come and go without prior notification and finding the right substitute for a given service is very often a hard task to achieve. In the literature, we distinguish three types of service similarity: those concerning the structural part or functional property of services, those concerning the behavioural part of services and those that deal with the non-functional properties of services. 

Structural similarity between services (\cite{12}) is a functional matching algorithm between the interface WSDL descriptions of Web services. The algorithm takes the description of Web services and is able to tell if the two services are similar using a semantic similarity metric. But this work need to be optimised and especially the non-functional parts of services need to be taken into account. Perse (\cite{8}) proposes a QoS metric for Web services based on normalization functions but this metric is used to dynamically compose services together. Perse does not consider service substitution as a separate problem from service composition. Finally, EurekaBESERIAL (\cite{13}) proposes an algorithm that is capable of detecting all the incompatibilities between two interface behaviour for Web services and based on these incompatibilities it introduces a similarity function to compare two Web service behaviour but it does not take into account the non-functional properties of services. 

Some works deal with service substitution (\cite{7,11}). Siroco (\cite{7})  proposes a framework that substitutes stateful Web services, taking into account the state of a service when executing and ensuring to applications a service continuity when substituting the service. But Siroco does not deal for now with non-functional properties. Santhanam, Basu, and Honavar (2009) propose a Web service substitution based on preferences over non-functional attributes but their description of non-functional properties is not general enough to take all types of non-functional properties into consideration. In this article, we do not limit our model to Web services as all major systems do but propose a general model of a service, describing its functional and non-functional properties (quantitative or qualitative) and based on this model we propose different metrics for computing non-functional service similarities. Than, we propose a mechanism for service substitution that substitute services not only upon their unavailability as the major systems do, but also when a new service fits better an application.

\section{Service Functional and Non-Functional QoS Equivalence Relations}

\subsection{Service Model}

We define a generic service model as composed of a functional interface and non-functional QoS properties. A functional interface specifies operations that can be performed on the service. An operation is described by a concept, a set of inputs and an output. The QoS non-functional properties describe the operation capabilities. These capabilities reflect the quality of the functionality expected from the service, such as dependability (including availability, reliability, security and safety), accuracy of the operation, speed of the operation, and so on. The service is also semantically described. The semantic description is upon the operations and QoS properties and is based upon common ontology concepts.

Consider finite sets of grammatical alphabet $\Sigma$, ontologies $\texttt{O}$, concepts $\texttt{N}$ belongings to these ontologies $\texttt{O}$, operations $\texttt{Op}$, inputs $\texttt{In}$, outputs $\texttt{Out}$, concepts $\texttt{Cpt}$, non-functional properties $\texttt{Np}$, quantitative and qualitative non-functional properties $\texttt{Np$_{QN}$}$, $\texttt{Np$_{QL}$}$. Consider the following operators: $^{*}$ (repetition zero or more times), $^{+}$ (repetition one or more times), $|\  |$ (the number of occurrences) and $^{0..1}$ (repetition zero or one time).
\\
\\
We define an operation $op$ belonging to $Op\ \subset\ \texttt{Op}$ as follows:
\\
\\
($op\ \in  Op\ \Leftrightarrow\ \exists\ In\ \subset \texttt{In},\ \exists\ Out\ \subset \texttt{Out},\ \exists\ cpt\ \in \texttt{Cpt},\ \exists\ Np\ \subset \texttt{Np},\ \exists\ Np_{QN}\ \subset \texttt{$Np_{QN}$},\ \exists\ Np_{QL}\ \subset \texttt{$Np_{QL}$}$):
\\
\\
$\begin{array}{llll}
op:\  <In^{*},\ Out^{0..1},\ cpt,\ Np^{*}> & \\\\
in:\ <name, type, semantic>,\ name\ \in\ \Sigma^{*}  & \\
out:\ <type, semantic> & \\
cpt:\ <name, semantic>,\ name\ \in\ \Sigma^{*}  & \\
type:\ <language, name>,\ \left\{name,\ language\right\} \in\ \Sigma^{*}  & \\
semantic:\ <o, n>,\ o\ \in\ O\ \subset \texttt{O} ,\ n\ \in\ N\ \subset \texttt{N}& \\
np:\ <Np^{*}_{QL}, Np^{*}_{QN}> & \\
np_{QL}:\ <name, semantic>,\ name\ \in\ \Sigma^{*}  & \\
np_{QN}:\ <name, numericValue, operator>,\ name\ \in\ \Sigma^{*} & \\
numericValue \in\ \mathbb{R} & \\
operator:\ \left\{<,\ >,\ \leq,\ \geq\right\} & \\
\end{array}$

where:
\begin{itemize}
	\item $In$ is the set of the operation $op$ inputs. $in$ is defined as a tuple where $name$ is the chosen input syntactic name, $type$ is the syntactic input type, and $semantic$ the input semantic description.
	\item $out\ \in\ Out$ is the operation $op$ output. $out$ is defined as a tuple where $type$ is the output syntactic type, and $semantic$ its semantic description. 
	\item $cpt$ is the concept the operation $op$ defines. The operation $op$ concept $cpt$ is defined as a tuple, where $name$ is the syntactic name through which the operation is called and $semantic$ its semantic description.
	\item $Np$ is the set of non-functional properties characterizing $op$. $Np$ can be qualitative or quantitative. $np_{QN}\ \in\ Np_{QL}$ is the qualitative non-functional properties defined as a tuple $<name, semantic>$. $np_{QN}\ \in\ Np_{QN}$ is the quantitative non-functional properties defined as a tuple, where $numericValue\ \in\ \R$ and $operator\ \in\ \left\{>,\ <,\ \leq,\ \geq\right\}$. $operator$ specifies the order applied to $numericValue$. For $\left\{>,\ \geq\right\}$ the greater the $numericValue$ is, the best is the QoS property for the service runtime execution. For $\left\{<,\ \leq\right\}$ the smaller the $numericValue$ is, the best is the QoS property for the service runtime execution.
\end{itemize}
The $type$ depends strongly on the programming language the $op$ is defined in, whereas the $semantic$ is independent of the technology and more related to the set of defined ontologies $O$.

Our service model is general enough to respect the SOA specifications, and to offer a common model to the heterogeneous technologies usually available in pervasive environments. The model proposes semantic descriptions relying on common ontologies, and by that it allows to abstract from the programming languages.

\begin{exple}

We consider three operations (cf. figure~\ref{figure_example-1}) and three interfaces (cf. figure~\ref{figure_ex-3}) described under the generic service model.
Each operation has a set of inputs described by a name, a type, and a semantic description, an output described by a type and a semantic description, and a concept described by a name and a semantic concept. Each operation can have one or several non-functional properties, qualitative or quantitative. These three operations (cf. figure~\ref{figure_example-1}) and three interfaces (cf. figure~\ref{figure_ex-3})  are used in the following examples to illustrate the upcoming definitions.

\begin{figure}[!ht]
\centering
\includegraphics[width=0.55\textwidth]{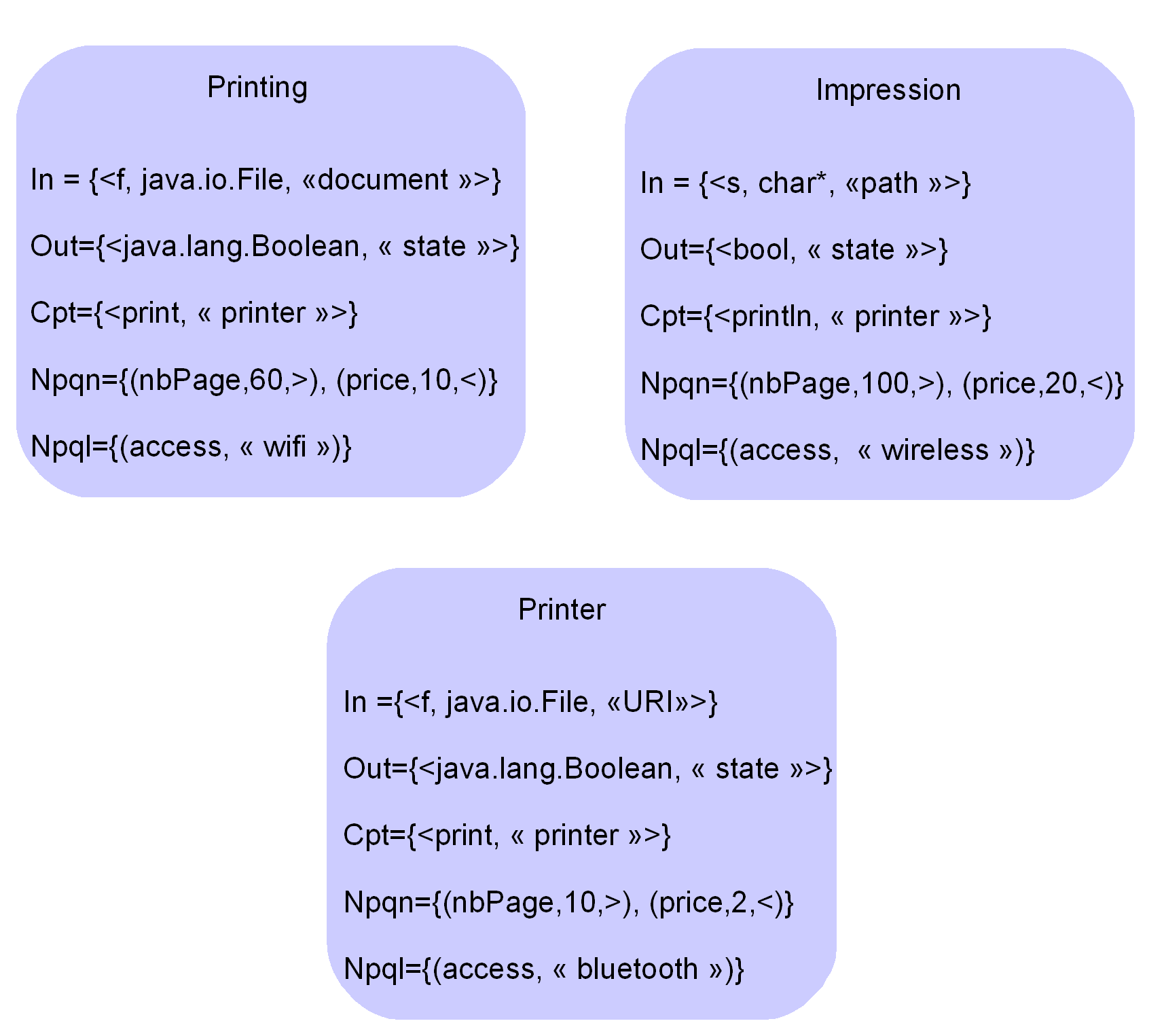}
\caption{Three operation specifications}
\label{figure_example-1}
\end{figure}

\begin{figure}[!h]
\centering
\includegraphics[width=0.65\textwidth]{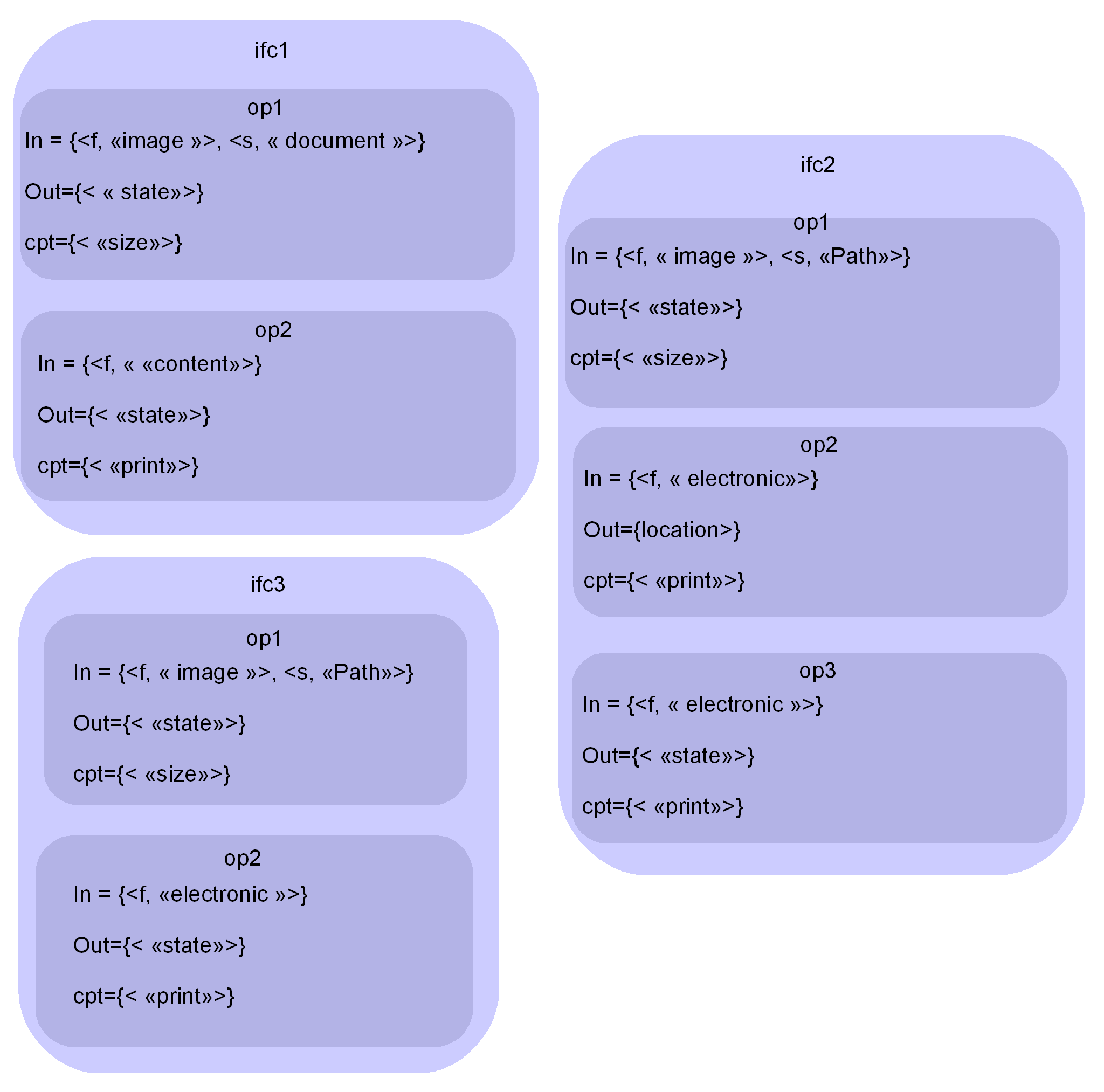}
\caption{Three interface specifications}
\label{figure_ex-3}
\end{figure}

\end{exple}

\subsection{Service Equivalence Relations}

\begin{figure}[!ht]
\centering
\includegraphics[width=0.4\textwidth]{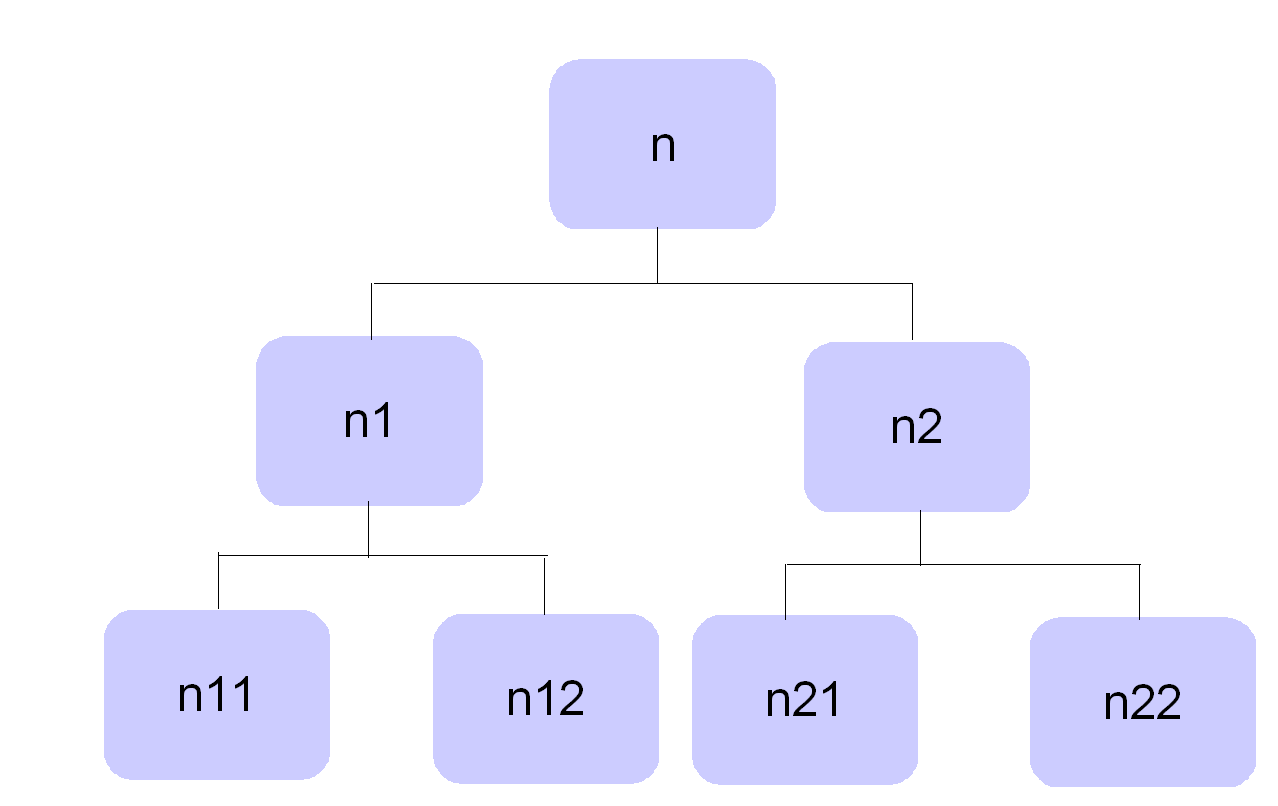}
\caption{An ontology example}
\label{figureontology1}
\end{figure}

Service equivalence relations determine whether two services offer the same functionality or not. A service is considered equivalent to another one if it can offer the same functionality (same interface) even with different non-functional QoS properties. The aim of this section is to provide definitions of possible relations between services in order to identify and decide when a service can be replaced by another one. Two relations are introduced: the equivalence ($\equiv$) and the almost equivalence ($\triangleright$) relations. In an equivalence relation, the two equivalent entities can interchange and be replaced one by the other. The equivalence relation is reflexive, symmetric, and transitive. In an almost equivalence relation only one entity can replace the other one. This relation is non reflexive, asymmetric, and transitive. It is based on sub-concept relations in the ontologies used to describe services of the environments. The relations tackle two main parts of a service: its functional interface and its non functional QoS properties. In the rest of this section, we define our interface equivalence relations and our non-functional QoS equivalence degree.
\\
\\
We define the interface equivalence $\equiv_{sem}$ upon the operation equivalence which itself is defined upon a concept matching $M_{Cpt}$ with concepts belonging to a defined ontology. We begin by defining the concept matching of a given ontology.
\\
\\
\noindent
\textbf{Concept matching}
\\
\\
The matching of two concepts belonging to the same ontology has been widely studied. We define our matching relation $M_{Cpt}$ between concepts belonging to the same ontology. A concept $n$ belonging to an ontology $o$ (figure~\ref{figureontology1}), can provide all its immediate sub-concepts $n_{1}$ and $n_{2}$ or one of its sub-concepts $n_{1}$ or $n_{2}$. This distinction depends strongly on the ontology definitions and providers. Some research such as Paolucci (\cite{3}) made the assumption that by selecting a concept $n$, we implicitly suppose that it provides all its immediate sub-concepts, others made the other assumption that by selecting a concept $n$, it provides at least one of its immediate sub-concepts, but not necessarily all of them.
Consider the set $\left\{n_{1}, n_{2}, .., n_{n}\right\}$ of all the sub-concepts of a concept $n$ in an ontology $o$, the assumption of Paolucci (\cite{3}) is formalised as follows:
$n \ \equiv_{provide} \ (n_{1} \wedge n_{2} \wedge ... \wedge n_{n})$ which means that $n$ can replace $n1$, $n2$, etc.
Others, do not make strong assumptions as this and suppose that a concept $n$ provides one or more of its sub-concepts but not necessarily all of them, $n \ \equiv_{provide} \ (n_{1} \vee n_{2} \vee ... \vee n_{n})$.
We fall into the first category, stipulating that a super-concept offers what its sub-concepts offer, and hence can replace them.

Defining $n$ and $m$, two concepts belonging to the same ontology $o$. We define the four values of concept matching $M_{Cpt}$ inspired from Paolucci (\cite{3}) as follows:

\begin{defi}{$M_{Cpt}(n,m)\ =\ Exact$}
If \ n\  and \ m\  are\  equivalent\  concept
\end{defi}

\begin{defi}{$M_{Cpt}(n,m)\ =\ PlugIn$}
If\  n\  is\  a\  super$-$concept\  of\  m
\end{defi}

\begin{defi}{$M_{Cpt}(n,m)\ =\ Subsume$}
If\  n\  is\  a\  sub$-$concept\  of\  m 
\end{defi}

\begin{defi}{$M_{Cpt}(n,m)\ =\ Fail$}
If\  n\  and\  m\  do\  not\  verify\  the\ above\ conditions
\end{defi}

\begin{exple}
Using our ontology example figure~\ref{figureontology}, we give an example of $M_{Cpt}$. 
\\
\\
$\begin{array}{l|lll}
Example & M_{Cpt} (''content'',\ ''electronic'')  = PlugIn && \\
& M_{Cpt} (''document'',\ ''URL'')  = PlugIn & & \\
& M_{Cpt} (''paper'',\ ''document'')  = Subsume && \\
& M_{Cpt} (''content'',\ ''path'')  = Fail && \\
\end{array}$

\begin{figure}[!h]
\centering
\includegraphics[width=0.45\textwidth]{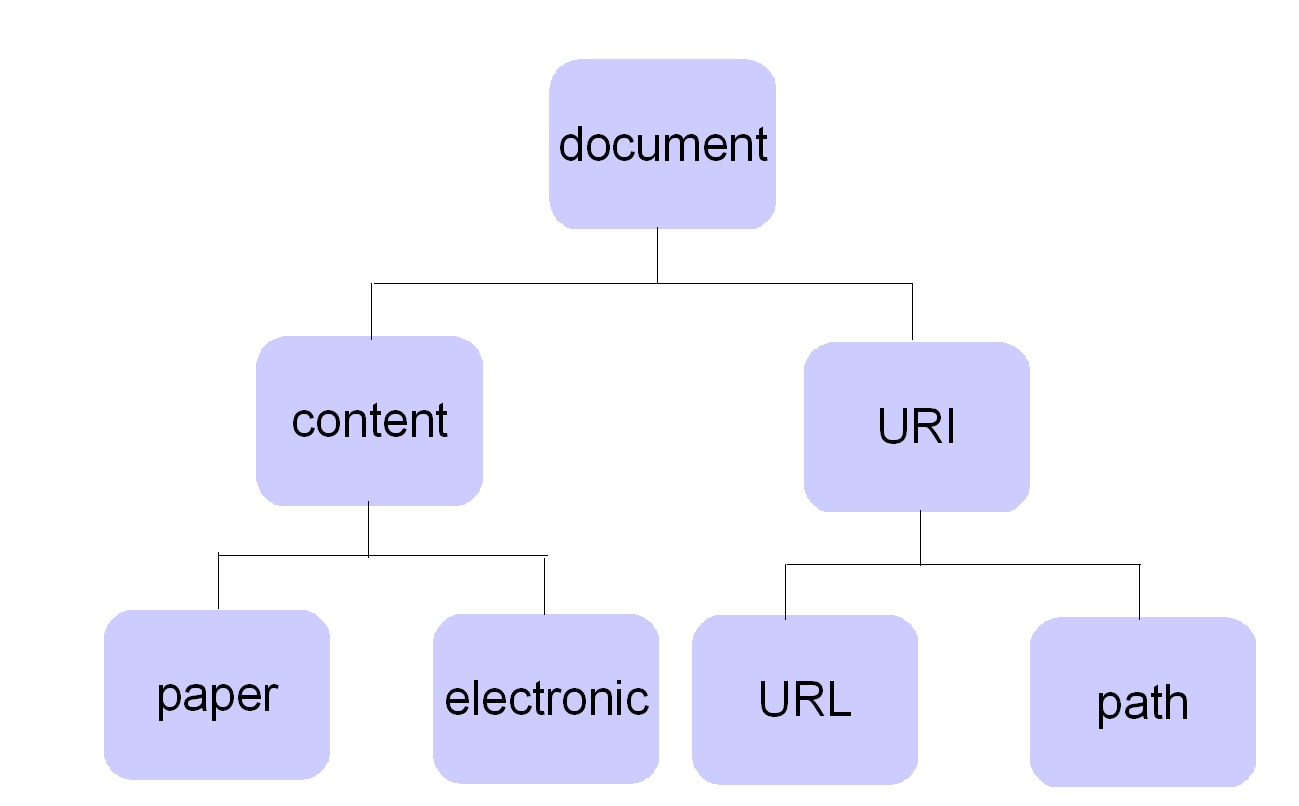}
\caption{A document ontology example}
\label{figureontology}
\end{figure}

\end{exple}

These concept matching values are the metrics employed to match operations and interfaces of services. We first define the values that the matching of operations can take, and based on these values we define when two operations are equivalent or almost equivalent.
\\
\\
\noindent
\textbf{Semantic operation equivalence}

\begin{defi}{Comparable Operations $\propto$}

We define two operations $opi$ and $opj$ to be comparable ($\propto(opi,opj)\ =\ true$) if they have the same number of inputs and the same number of outputs and if it exists a bijection $f$ over their inputs allowing to compare the inputs parameters two by two. $\forall\ k,l\ \in\ \left\{1..|In_{opi}|\right\}$
\\
\\
$\begin{array}{llll}
& |In_{opi}|\ =\ |In_{opj}| \wedge\ (|Out_{opi}|\ =\ |Out_{opj}|)  & \\
& \wedge\ (\exists\ f: In_{opi}\ \rightarrow\ In_{opj},\ \forall\ in_{l}\ \in\ In_{opj},\ \exists!\ in_{k}\ \in\ In_{opi},\ f(in_{k})= in_{l}) && 
\end{array}$
\end{defi}

$\forall\ \left\{i,j,l,k\right\} \in \N$, we define the semantic matching, $M_{sem}(op_{i},op_{j})$, of two comparable operations $op_{i}$ and $op_{j}$ ($\propto(op_{i},op_{j})\ =\ true$), considering the semantic matching of their concepts, inputs and outputs.

We can quickly realize that the semantic matching of these three items - inputs, outputs, and concepts - can be different, as the concept matching can take multiple values. In a semantic matching, the three items can range from $Exact$ matching to $Fail$ passing by the $PlugIn$ and $Subsume$ values.
\\
\\
We define the different values a semantic matching $M_{sem}$ between two operations $op_{i}$ and $op_{j}$ can take as follows:
\\
\\
\begin{defi}{$ M_{sem}(op_{i},op_{j})\ =\ Exact$}
Two operations $op_{i}$ and $op_{j}$ verifying ($\propto(op_{i},op_{j})\ =\ true$) are $Exact$ semantic matching if all the matching values between concept, inputs and output are $Exact$. $\forall\  k \in \N$:
\\
$\begin{array}{llll}
& (M_{Cpt}(sem_{cpt_{op_{i}}},  sem_{cpt_{op_{j}}}) =\ Exact) & \\
& \wedge\ (\forall\  in_{k}\ \in\ In_{op_{i}},\ M_{Cpt}(sem_{in_{k}}, f(sem_{in_{k}}))\ =\ Exact) && \\
& \wedge\ (M_{Cpt}(sem_{out_{op_{i}}},sem_{out_{op_{j}}}) =\ Exact)
\end{array}$
\end{defi}

\begin{defi}{$ M_{sem}(op_{i},op_{j})\ =\ PlugIn$}
They are $PlugIn$ semantic matching if they are not $Exact$ matching and all the matching between concept, inputs or output values are $Exact$ or $PlugIn$. $\forall\ k \in \N$:
\\
$\begin{array}{llll}
& M_{sem}(op_{i},op_{j})\ \neq\ Exact && \\
& \wedge\ (M_{Cpt}(sem_{cpt_{op_{i}}},  sem_{cpt_{op_{j}}}) \in\ \left\{Exact\ \vee\ PlugIn\right\}) & \\
& \wedge\ (\forall\  in_{k}\ \in\ In_{op_{i}},\ M_{Cpt}(sem_{in_{k}}, f(sem_{in_{k}}))\ in\ \left\{Exact \vee PlugIn\right\}) && \\
& \wedge\ (M_{Cpt}(sem_{out_{op_{i}}},sem_{out_{op_{j}}}) \in\ \left\{Exact \vee PlugIn\right\})
\end{array}$
\end{defi}

\begin{defi}{$ M_{sem}(op_{i},op_{j})\ =\ Subsume$}
They are $Subsume$ semantic matching if  they are no $Exact$ or $PlugIn$ matching and at least one matching value between concept, inputs or output is $Subsume$ and no $Fail$ matching value is found between outputs, concepts, and the corresponding comparable inputs. $\forall\ k \in \N$:
\\
$\begin{array}{llll}
& M_{sem}(op_{i},op_{j})\ \neq\ Exact && \\
& \wedge\ (M_{sem}(op_{i},op_{j})\ \neq\ PlugIn) && \\
& \wedge\ (M_{Cpt}(sem_{cpt_{op_{i}}},  sem_{cpt_{op_{j}}}) =\ \neg(Fail)) & \\
& \wedge\ (\forall\  in_{k}\ \in\ In_{op_{i}},\ M_{Cpt}(sem_{in_{k}}, f(sem_{in_{k}}))\ =\ \neg(Fail))&& \\
& \wedge\ (M_{Cpt}(sem_{out_{op_{i}}},sem_{out_{op_{j}}}) =\ \neg(Fail))
\end{array}$
\end{defi}

\begin{defi}{$ M_{sem}(op_{i},op_{j})\ =\ Fail$}
They are $Fail$ semantic matching if they have different inputs or outputs numbers or at least one semantic matching value between concepts, inputs or outputs is $Fail$. $\forall\ \left\{k,l\right\} \in \N$:
\\
$\begin{array}{llll}
& (|In_{op_{i}}|\ \neq\ |In_{op_{j}}|) & &\\
& \vee\ (|Out_{op_{i}}|\ \neq\ |Out_{op_{j}}|) & &\\
& \vee\ (M_{Cpt}(sem_{cpt_{op_{i}}},  sem_{cpt_{op_{j}}}) =\ Fail) & \\
& \vee\ (\exists\  in_{k}\ \in\ In_{op_{i}},\ \forall\ in_{l}\ \in\ In_{op_{j}},\ M_{Cpt}(sem_{in_{k}}, sem_{in_{l}})\ =\ Fail) && \\
& \vee (M_{Cpt}(sem_{out_{op_{i}}},sem_{out_{op_{j}}}) =\ Fail)
\end{array}$
\end{defi}


\begin{exple}
Considering the three operations defined in figure~\ref{figure_example-1}

The semantic matching between these operations give the following values:
\\
\\
$\begin{array}{llll}
& M_{Cpt}(Printing, Impression)\ =\ PlugIn & &\\
& M_{Cpt}(Printing, Printer)\ =\ PlugIn & &\\
& M_{Cpt}(Impression, Printer)\ =\ Subsume & \\
& M_{Cpt}(Impression, Printing)\ =\ Subsume && \\
& M_{Cpt}(Printer, Printing)\ =\ Subsume && \\
& M_{Cpt}(Printer, Impression)\ =\ PlugIn &&
\end{array}$

\end{exple}

The semantic operation matching provides the tools to define when operations are equivalent or almost equivalent.

\begin{defi}{Operation equivalence}
We define two operations $op_{i}$ and $op_{j}$ to be semantically equivalent $\equiv_{sem}$ if:
\\
\\
$(\equiv_{sem}(op_{i},op_{j})\ =\ true)\ \Leftrightarrow\ (M_{sem}(opi,opj)\ =\ Exact)$
\\
\\
The operation equivalence $\equiv_{sem}$ is reflexive, symmetric, and transitive. We notify that the semantic equivalence satisfies the conditions an equivalence relation $\Re$ needs to fulfill.
\end{defi}

\begin{defi}{Operation almost equivalence}
We define two operations $op_{i}$ and $op_{j}$ to be semantically almost equivalent $\triangleright_{sem}$ if:
\\
\\
$(\triangleright_{sem}(opi,opj)\ =\ true)\ \Leftrightarrow\ (M_{sem}(opi,opj)\ =\ PlugIn)$
\\
\\
The almost equivalence is non reflexive, asymmetric, and transitive. This relation of almost equivalence specifies that $opi$ is equivalent to $opj$ and can replace it but that the contrary is not true. $opj$ can not always replace $opi$.
\end{defi}

\begin{exple}
Coming back to our example in figure~\ref{figure_example-1}, where we had these matching values between the three operations $Printing$, $Impression$, and $Printer$:
\\
\\
$\begin{array}{llll}
& M_{Cpt}(Printing, Impression)\ =\ PlugIn & &\\
& M_{Cpt}(Printing, Printer)\ =\ PlugIn & &\\
& M_{Cpt}(Impression, Printer)\ =\ Subsume & \\
& M_{Cpt}(Impression, Printing)\ =\ Subsume && \\
& M_{Cpt}(Printer, Printing)\ =\ Subsume && \\
& M_{Cpt}(Printer, Impression)\ =\ PlugIn &&
\end{array}$
\\
\\
we can conclude the following almost equivalent relations:
\\
\\
$\begin{array}{llll}
& \triangleright(Printing, Impression)\ =\ true & &\\
& \triangleright(Printing, Printer)\ =\ true & &\\
& \triangleright(Printer, Impression)\ =\ true &&
\end{array}$
\end{exple}

Now that we have defined the operation equivalence relations, we define the interface equivalence relations and by that we define when two services are equivalent or almost equivalent.
\\
\\
\noindent
\textbf{\small{Interface equivalence}}
\\
\\
We define two interfaces to be comparable ($\propto(ifc_{i},ifc_{j})\ =\ true$) if they have the same number of operations and if it exists a bijection $f$ over their operations allowing to compare them two by two:

\begin{defi}{Comparable interfaces $\propto$}

We define two interfaces $ifc_{i}$ and $ifc_{j}$ to be comparable ($\propto(ifc_{i},ifc_{j})\ =\ true$) if:
\\
\\
$\begin{array}{llll}
& |Op_{ifc_{i}}|\ =\ |Op_{ifc_{j}}| & \\
& \wedge\ (\exists\ f: Op_{ifc_{i}}\ \rightarrow\ Op_{ifc_{j}},\ \forall\ op_{l}\ \in\ Op_{ifc_{j}},\ \exists!\ op_{k}\ \in\ Op_{ifc_{i}},\ f(op_{k})= op_{l}) && 
\end{array}$
\end{defi}

As for operations we define the semantic matching between two interfaces $ifc_{i}$ and $ifc_{j}$:
\\
\begin{defi}{$M_{sem}(ifc_{i},ifc_{j})\ =\ Exact$}
Two interfaces $ifc_{i}$ and $ifc_{j}$ are $Exact$ semantic match if $\propto(ifc_{i},ifc_{j})\ =\ true$ and:
\\
\\
$\begin{array}{llll}
& \forall op_{i}\ \in\ Op_{ifc_{i}},\ M_{sem}(op_{i}, f(op_{i}))\ =\ Exact &&
\end{array}$
\end{defi}

\begin{defi}{$M_{sem}(ifc_{i},ifc_{j})\ =\ PlugIn$}
They are $PlugIn$ semantic match if  $\propto(ifc_{i},ifc_{j})\ =\ true$, and:
\\
\\
$\begin{array}{llll}
& M_{sem}(ifc_{i},ifc_{j})\ \neq\ Exact && \\
& \wedge\ (\forall op_{i}\ \in\ Op_{ifc_{i}},\ M_{sem}(op_{i}, f(op_{i}))\ \in\ \left\{Exact \vee\ PlugIn\right\}) &&
\end{array}$
\end{defi}

\begin{defi}{$M_{sem}(ifc_{i},ifc_{j})\ =\ Subsume$}
They are $Subsume$ semantic match if $\propto(ifc_{i},ifc_{j})\ =\ true$, $ifc_{i}$ and $ifc_{j}$ are not $Exact$ nor $PlugIn$ semantic match and:
\\
\\
$\begin{array}{llll}
& M_{sem}(ifc_{i},ifc_{j})\ \neq\ Exact && \\
& \wedge\ (M_{sem}(ifc_{i},ifc_{j})\ \neq\ PlugIn) && \\
& \wedge\ (\forall op_{i}\ \in\ Op_{ifc_{i}},\ M_{sem}(op_{i}, f(op_{i}))\ \in\ \left\{Exact \vee\ PlugIn\ \vee\ Subsume\right\}) &&
\end{array}$
\end{defi}

\begin{defi}{$M_{sem}(ifc_{i},ifc_{j})\ =\ Fail$}
They are $Fail$ semantic match if:
\\
\\
$\begin{array}{llll}
& \propto(ifc_{i},ifc_{j})\ =\ false && \\
& \vee\ (\exists\  op_{i}\ \in\ Op_{ifc_{i}},\ \forall\  op_{j}\ \in\ Op_{ifc_{j}},\ M_{sem}(op_{i},op_{j})\ =\ Fail) &&
\end{array}$
\\
\\
It is sufficient to have only one operation $op_{i}$ of $ifc_{i}$ that do $Fail$ match with any operation $op_{j}$ of $ifc_{j}$ to declare that the two services matching fails.
\end{defi}

Based on these interface semantic matching definitions, we define the interface equivalence and almost equivalence.
\begin{defi}{Interface equivalence}
We define two interfaces $ifc_{i}$ and $ifc_{j}$ to be semantically equivalent $\equiv_{sem}$ if:
\\
\\
$(\equiv_{sem}(ifc_{i},ifc_{j})\ =\ true)\ \Leftrightarrow\ (M_{sem}(ifc_{i},ifc_{j})\ =\ Exact)$
\\
\\
The equivalence $\equiv_{sem}$ is reflexive, symmetric, and transitive.
\end{defi}

\begin{defi}{Interface almost equivalence}
We define two services $ifc_{i}$ and $ifc_{j}$ to be semantically almost equivalent $\triangleright_{sem}$ if:
\\
\\
$(\triangleright_{sem}(ifc_{i},ifc_{j})\ =\ true)\ \Leftrightarrow\ (M_{sem}(ifc_{i},ifc_{j})\ =\ PlugIn)$
\\
\\
As for operations, the almost equivalence is non reflexive, non symmetric, and transitive. This relation of almost equivalence specifies that $ifc_{i}$ is equivalent to $ifc_{j}$ and can replace it but that the contrary is not true. $ifc_{j}$ cannot always replace $ifc_{i}$.
\end{defi}

\begin{exple}
Considering the three interfaces and their semantic descriptions in figure~\ref{figure_ex-3}:

The semantic matching between their different operations gives the following values:
\\
\\
$\begin{array}{llll}
& \propto(ifc1,ifc3)\ =\ true  & \\
& \wedge\ (M_{sem}(op1_{ifc1}, op1_{ifc3})\ =\ PlugIn) & &\\
& \wedge\ (M_{sem}(op2_{ifc1}, op2_{ifc3})\ =\ PlugIn) &
\end{array}$
\\
\\
We can implies $\ \Rightarrow\ (\triangleright_{sem}(ifc1,ifc3)\ =\ true)$
\\
\\
The two interfaces $ifc1$ and $ifc2$ are not comparable as they do not have the same number of operations. Nevertheless, some of their operations are $PlugIn$ semantic.
\end{exple}

Many services do not have the same number of operations per interface as depicted in example 5. To resolve this issue brought by the example. We define the matching over a set of operations for two interfaces $ifc_{i}$ and $ifc_{j}$.

\begin{defi}{$M^{Op}_{sem}(ifc_{i},ifc_{j})\ =\ Exact $}
Two interfaces $ifc_{i}$ and $ifc_{j}$ are $Exact$ semantic matching over a subset of operations $Op$, if:
\\
\\
$\begin{array}{llll}
& \propto^{Op}(ifc_{i},ifc_{j})\ =\ true && \\
& \wedge\ (Op\ \subset\ Op_{ifc_{i}},\ \forall opi\ \in\ Op,\ M_{sem}(op_{i}, f(op_{i}))\ =\ Exact) &&
\end{array}$
\end{defi}

\begin{defi}{$M^{Op}_{sem}(ifc_{i},ifc_{j})\ =\ PlugIn$}
Two services $ifc_{i}$ and $ifc_{j}$ are $PlugIn$ semantic matching over a subset of operations $Op$, if:
\\
\\
$\begin{array}{llll}
& \propto^{Op}(ifc_{i},ifc_{j})\ =\ true && \\
& M^{Op}_{sem}(ifc_{i},ifc_{j})\ \neq\ Exact && \\
& \wedge\ (Op\ \subset\ Op_{ifc_{i}},\ \forall opi\ \in\ Op,\ M_{sem}(op_{i}, f(op_{i}))\ \in\ \left\{Exact \vee\ PlugIn\right\}) &&
\end{array}$
\end{defi}

We thus define interface equivalence and almost equivalence between interfaces over a subset of operations:
\\
\\
\begin{defi}{Interface equivalence over a subset of operations, $\equiv^{Op}_{sem}$}
We define two interfaces $ifc_{i}$ and $ifc_{j}$ to be semantically equivalent over a subset of operations $Op$:
\\
\\
$(\equiv^{Op}_{sem}(ifc_{i},ifc_{j})\ =\ true)\ \Leftrightarrow\ (M^{Op}_{sem}(ifc_{i},ifc_{j})\ =\ Exact)$
\\
\\
\end{defi}

\begin{defi}{Interface almost equivalence over a subset of operations, $\triangleright^{Op}_{sem}$}
We define two services $ifc_{i}$ and $ifc_{j}$ to be semantically almost equivalent over a subset of equivalence $Op$:
\\
\\
$(\triangleright^{Op}_{sem}(ifc_{i},ifc_{j})\ =\ true)\ \Leftrightarrow\ (M^{Op}_{sem}(ifc_{i},ifc_{j})\ =\ PlugIn)$
\\
\\
\end{defi}

\begin{exple}
Coming back to our example in figure~\ref{figure_ex-3}. The semantic matching between the different operations of $ifc1$ and $ifc2$ gives the following values:
\\
\\
$\begin{array}{llll}
& \propto^{op1_{ifc1}, op2_{ifc1}}(ifc1,ifc2)\ =\ true  & \\
& \wedge\ (M_{sem}(op1_{ifc1}, op1_{ifc2})\ =\ PlugIn) & &\\
& \wedge\ (M_{sem}(op2_{ifc1}, op3_{ifc2})\ =\ PlugIn) & &
\end{array}$
\\
\\
\\
From these matching values, we can implies $\ \Rightarrow\ (\triangleright^{\left\{op1_{ifc1},\ op2_{ifc1}\right\}}_{sem}(ifc1,ifc2)\ =\ true)$
\\
\\
The two interfaces $ifc1$ and $ifc2$ are almost equivalent upon the two operations of $ifc1$.
\end{exple}

This equivalence and almost equivalence over subsets of operations is useful for service substitution issues, as a service can be replaced by another one if certain operations are specified to be required by applications at a given time. 
\\
Many services can be almost equivalent and we need to be able to rank between these almost equivalence relations. A ranking of the semantic matching values need to be introduced. This ranking will help ordering services that have semantic almost equivalence with different concept values for the respective operations' inputs, outputs and concepts. It is also used to rank interfaces and operations that have $Subsume$ semantic matching. This operations' ordering allows users and applications to choose services that best suit their requirements at a given time, and re-adapt their choice if other services that have a closer semantic equivalence appear. We introduce a semantic distance $D_{sem}$ between two interfaces. It calculates the distance between two interfaces semantic descriptions. The more this value is closer to zero the more these two services are equivalent.
\\
\\
\noindent
\textbf{Semantic distance}
\\
\begin{defi}{Concept semantic distance}
We first define a normalised concept distance $D_{Cpt}$ between two concepts $n$ and $m$: 
\\
\\
$\begin{array}{llll}
D_{Cpt}(n,m):& 0\ \hspace{1cm}  if\ M_{Cpt}(n,m)\  = Exact && \\
& 0.2\ \hspace{0.7cm} if\ M_{Cpt}(n,m)\  = PlugIn & & \\
& 0.8\ \hspace{0.7cm}  if\ M_{Cpt}(n,m)\  = Subsume && \\
& 1\ \hspace{1cm}  if\ M_{Cpt}(n,m)\  = Fail && 
\end{array}$
\\
\\
The closer the distance is to zero, the best is the semantic value matching between two concepts. An $Exact$ value is preferred to a $PlugIn$ one, which is preferred to a $Subsume$ one. The choice of values can vary. The idea is to assign different values and especially values that reflect the importance of the matching result. In this definition we chose to distinguish to Exact and PlugIn from Subsume and Fail. Other values more ponderated can be chosen.
\end{defi}

\begin{defi}{Operation semantic distance}
We define the semantic distance between two comparable operations $opi$ and $opj$ ($\propto(opi,opj)\ =\ true$): ($D_{sem}(opi,opj),\ i,j\ \in \N$). This semantic distance is the sum of the ponderated concept distance of the operation concept, inputs and output semantic description:
\\
\\
$w_{1}*D_{Cpt}(sem_{cpt{opi}}, sem_{cpt{opj}}) + w_{2}*D_{Cpt}(sem_{out_{opi}}, sem_{out_{opj}}) + \\ \sum^{|In_{opi}|}_{k=1}(w_{k}*D_{Cpt}(sem_{ink_{opi}}, sem_{f(ink_{opi})}))$
\end{defi}

where $\sum_{i\in \N}(wi)\ =\ 1$
\\
$wi$ corresponds to the weight we wish to give to the concept, inputs and output. When matching two operations, the focus may be put on inputs, outputs parameters or on the concept. $wi$ allows to ponderate the ranking of operations.

\begin{defi}{Interface Semantic Distance}
The semantic distance between two comparable interfaces ($D_{sem}(ifc_{i},ifc_{i})\ i,j\ \in \N$) is the sum of all the semantic distance between their comparable operations, ponderated by a weight allowing to focus on some operations rather than others.
\\
\\
$\begin{array}{llll}
& \sum^{|Op_{ifc_{i}}|}_{k=1}(w_{k}*D_{sem}(opk_{ifc_{i}}, f(opk_{ifc_{i}}))) &&
\end{array}$
\end{defi}

\begin{exple}
We come back to our example and calculate the semantic distance $D_{sem}$ of our three operations:
\\
\\
$\begin{array}{l|lll}
&  D_{Cpt}(''printer'',''printer'')  = 0 && \\
&  D_{Cpt}(''document'',''URI'')  = 0.2 & & \\
&  D_{Cpt}(''state'',''state'')  = 0 && 
\end{array}$
\\
\\
$=>\ D_{sem}(printing, printer)\ =\ w2*0.2 $
\\
\\
$\begin{array}{l|lll}
 &  D_{Cpt}(''printer'',''printer'')  = 0 && \\
&  D_{Cpt}(''document'',''path'')  = 0.2 & & \\
&  D_{Cpt}(''state'',''state'')  = 0 && 
\end{array}$
\\
$=>\ D_{sem}(printing, impression)\ =\ w2* 0.2$
\\
\\
The operations $Printing$ is PlugIn matching with $Impression$ and $Printer$ and has the same semantic distance value to both operations.
\\
\\
$\begin{array}{l|lll}
& D_{Cpt}(''printer'',''printer'')  = 0 && \\
& D_{Cpt}(''path'',''document'')  = 0.8 & & \\
& D_{Cpt}(''state'',''state'')  = 0 && 
\end{array}$
\\
\\
$=>\ D_{sem}(impression, printing)\ =\ w2*0.8 $
\\
\\
The overall value of $D_{Cpt}(printing, impression)\ <\ D_{Cpt}(impression, printing)$ and is normal as $printing$ is PlugIn of $Impression$ and $Impression$ Subsume of $Printing$ (see example 3).
\end{exple}

\begin{exple}
We come back to our example and calculate the semantic distance $D_{sem}$ of our interfaces:
\\
\\
$\begin{array}{l|lll}
&  D_{Cpt}(op1_{ifc1},op1_{ifc3})  = 0.2 && \\
&  D_{Cpt}(op2_{ifc1},op2_{ifc3})  = 0.2 & &
\end{array}$
\\
\\
$=>\ D_{sem}(ifc1, ifc3)\ =\ w1*0.2 + w2*0.2 $
\\
\\
$\begin{array}{l|lll}
&  D_{Cpt}(op1_{ifc1},op1_{ifc2})  = 0.2 && \\
&  D_{Cpt}(op2_{ifc1},op3_{ifc2})  = 0.2 & &
\end{array}$
\\
$=>\ D_{sem}(ifc1, ifc2)\ =\ w1*0.2 + w2*0.2$
\\
\\
The overall values of $D_{Cpt}(ifc1, ifc3)$ and $D_{Cpt}(ifc1, ifc2)$ depends on the values assigned to the weights (w1 and w2).
\end{exple}

In our semantic distance calculation example, we gave the three items of an operation - inputs, output, and concept - the same importance. We can ponderate the semantic distance by introducing weights to each of the operation items. 

The equivalences introduced so far concern the interfaces of services. If two services can publish the same interface, they can provide different non-functional properties. If we are able to distinguish services by their functionalities, it is interesting to evaluate how equivalent services are in terms of non-functional QoS properties.  In the following section, we define  a metric to calculate the non-functional QoS equivalence degree for the services that are equivalent and almost equivalent.

\subsection{Non-Functional QoS Equivalence Degree}

Services can be  semantically equivalent, almost equivalent, or having $Subsume$ matching relations. These equivalence are based on the functional aspect of services. Services can offer the same functionalities but with different non-functional QoS properties. We will define a metric that measures the non-functional QoS degree of equivalence. This metric allows to assign a normalised degree that measures the degree of non-functional QoS similarities between two equivalent, almost equivalent, or $Subsume$ matching services. These degrees are used to choose between diverse services providing different non-functional QoS properties, but offering similar functionalities. 
\\
\\
The non-functional QoS of an operation is defined as follows:

\begin{defi}{non-functional QoS properties}
Consider a finite set of grammatical alphabet $\Sigma$, ontologies $O$, concepts $N$ belongings to these ontologies $O$, non-functional QoS properties $Np$, quantitative non-functional properties $Np_{QN}$, and qualitative non-functional properties $Np_{QL}$. Considering an operation $op$ we define its non-functional QoS as follows:
\\
\\
$\begin{array}{llll}
& Np:\ \left\{Np^{*}_{QL}, Np^{*}_{QN}\right\}  & &\\
& Np_{QL} = \left\{np1_{QL},\ np2_{QL},\ ..\ npk_{QL}\right\},\ k\ =\ |NP_{QL}| & &\\
& Np_{QN} = \left\{np1_{QN},\ np2_{QN},\ ..\ npt_{QN}\right\},\ k\ =\ |NP_{QN}| & &\\
& np_{QL} = <name, semantic>,\ name\ \in\ \Sigma^{*} & & \\
& np_{QN} = <name, numericValue, operator>,\ name\ \in\ \Sigma^{*}\ \& \ numericValue\ \in\ \R &&\\
& operator = \left\{<,\ >,\ \leq,\ \geq\right\} & & \\
& semantic = <o, n>,\ o\ \in\ O,\ n\ \in\ N& &
\end{array}$

\end{defi}

$operator$ specifies the order applied to $numericValue$. For $\left\{>,\ \geq\right\}$ the greater the $numericValue$ is, the best is the QoS property for the service runtime execution. For $\left\{<,\ \leq\right\}$ the smaller the $numericValue$ is, the best is the QoS property for the service runtime execution.

The non-functional equivalence degree  $QoS_{Degree}(opi, opj)$ between two functional equivalent operations is evaluated upon their quantitative and qualitative properties similarities. Two functional equivalent operations offer the same functionality but not necessarily the same non-functional QoS properties. The \\$QoS_{Degree}(opi, opj)$ evaluates the degree of similarities of two operations $opi$ and $opj$ concerning their non-functional QoS properties. We suppose that:
\\
$\exists\ f:\ Np_{opi}\ \rightarrow\ Np_{opj}\ where\ \forall\ npk_{opj}\ \in\ Np_{opj},\ \exists!\ npk_{opi}\ \in\ Np_{opi},\ f(npk_{opi})=npk_{opj}$.
\\
$\forall\ k\ \in\ \N,\ npk_{opi}$ and $npk_{opj}$ deals with the same non-functional QoS property.
 If $npk_{opi}$ is a quantitative non-functional QoS we have $npk_{opj}$ also a quantitative non-functional QoS and $name_{npk_{opi}}\ =\ name_{npk_{opj}}$. 
 \\
 \noindent
 If $npk_{opi}$ is a qualitative non-functional QoS we have $npk_{opj}$ also a qualitative non-functional QoS and $name_{npk_{opi}}\ =\ name_{npk_{opj}}$.

\begin{defi}{$QoS_{Degree}(opi, opj)$}
Considering two operations $opi$ and $opj$, we define the degree of equivalence between the two operations $QoS_{Degree}(opi, opj)$ as a function that measures how close is $opj$ from $opi$ in terms of non-functional QoS. We consider the non-functional properties of $opi$, $NP_{opi}$ and calculate as follows the degree of equivalence $opj$ has upon these properties: 
\\
\\
$\begin{array}{llll}
& QoS_{Degree}(opi, opj) = \sum_{k=1}^{|Np_{opi}|}w_{k}\ast\ deg(npk_{opi},npk_{opj}) & & \\
& && \\
\end{array}$
\end{defi}

where, $w_{k}$ is the assigned weight for a particular non-functional QoS property with the following conditions $\sum^{|Np_{opi}|}_{k=1}(w_{k})\ = 1$. The more $w_{k}$ is closer to zero, the more important is the property $Npk$. This ponderation allows to decide when searching for equivalent services if certain non-functional QoS properties are more important than other for the required service replacement. $deg(npk_{opi},npk_{opj})$ are normalised values between $0$ and $1$ corresponding to the equivalence degree between $npk_{opi}$ and $npk_{opj}$. These values are calculated using the z-score or standardization of the $npk$ values for quantitative properties and semantic distance for qualitative properties.

We define $deg(npk_{opi},npk_{opj})$ as follows:

\begin{itemize}
	\item $deg(npk_{opi},npk_{opj}) = deg(npk_{QN_{opi}},npk_{QN_{opj}})$ for the quantitative properties.
	\item $deg(npk_{opi},npk_{opj}) = deg(npk_{QL_{opi}},npk_{QL_{opj}})$ for the qualitative ones.
\end{itemize}

We define next how we calculate these two degrees.

\begin{defi}{$deg(npk_{QN_{opi}},npk_{QN_{opj}})$}
$deg(npk_{QN_{opi}},npk_{QN_{opj}})\ =\ |\eta(npk_{QN_{opi}})\ -\ \eta(npk_{QN_{opj}})| $
\\
We define $\eta(npk_{QN})$ as the normalization of z-score value of $npk_{QN}$ for quantitative non-functional QoS.
\end{defi}

\begin{defi}{$\eta(np_{QN})$}
Considering $np_{QN} = <name, numericValue, operator>$ we define $\eta(np_{QN})$ as follows
\\
\\
$\begin{array}{llll}
if\ operator_{np_{QN}}\ is\ '<'\ : & 0\ if\ z$-$score(np_{QN})\ <\ -2 & & \\
& 1\ if\ z$-$score(np_{QN})\ >\ 2 & & \\
& (z-score(np_{QN}))/4\ +\ 0.5\ if\ 2\ >\ z$-$score(np_{QN})\ >\ -2 & &
\end{array}$
\\
\\
$\begin{array}{llll}
if\ operator_{np_{QN}}\ is\ '>'\ : & 1\ if\ z$-$score(np_{QN})\ <\ -2 & & \\
& 0\ if\ z$-$score(np_{QN})\ >\ 2 & & \\
& 0.5\ -\ (z$-$score(np_{QN}))/4\ if\ 2\ >\ z$-$score(np_{QN})\ >\ -2 & &
\end{array}$
\end{defi}

For $<$ the $numericValue$ is the best when it is the smallest. $\eta(np_{QN})$ is closer to zero for the smallest value of $numericValue$ and closer to one for the bigger value of $numericValue$, and vice versa for $>$.

The z-score of a quantitative property $np_{QN}$, indicates how far and in what direction, the property deviates from its distribution's mean, expressed in units of its distribution's standard deviation. We use the z-score standardization in order to provide a way of comparing all the different non-functional QoS by including consideration of their respective distributions.

\begin{defi}{$z$-$score(np_{QN})$}
Considering the quantitative $np_{QN}$, its corresponding z-score is:
\\
\\
$\begin{array}{llll}
z$-$score(np_{QN}) = & (numericValue_{np_{QN}}\ -\ \mu(numericValue_{np_{QN}}))/\sigma(numericValue_{np_{QN}}) & &
\end{array}$
\\
\\
where, $\mu(numericValue_{np_{QN}})$ is the mean of the values of $np_{QN}$, and $\sigma(numericValue_{np_{QN}})$ is the standard deviation of $np_{QN}$.
\end{defi}

In normal distribution we can distinguish that the $95\%$ of z-score($np_{QN}$) values are comprises between $-2$ and $2$. Based on this, $\eta(np_{QN})$ calculates a value between $0$ and $1$ taking into account the nature of quantitative non-functional QoS properties. Indeed the $operator_{np_{QN}}$ indicates whether the properties are stronger with greater values, or with smaller values.


If for the quantitative non-functional QoS properties, we used z-score and normalization to calculate the degree of similarities between two properties, for qualitative non-functional QoS we use the semantic distance to compare the concepts of the qualitative properties $np_{QL}$. The semantic distance returns a normalised value between $0$ and $1$.
\\
\begin{defi}{$deg(npk_{QL_{opi}},npk_{QL_{opj}})$}
Considering $npk_{QL_{opi}}$ the qualitative non-functional QoS of the operation. We seek to find the best equivalence for it from a set of equivalent operations.
Considering $npk_{QL_{opj}} = <name, semantic>$ the qualitative non-functional QoS of the other operations. we define $deg(npk_{QL_{opi}},npk_{QL_{opj}})$ as follows:
\\
\\
$\begin{array}{llll}
&deg(npk_{QL_{opi}},npk_{QL_{opj}})\ =\ D_{sem}(n_{semantic_{npk_{QL_{opi}}}}, n_{semantic_{npk_{QL_{opi}}}}) & & \\
\end{array}$
\end{defi}

\begin{exple}
Considering the three operations defined in figure~\ref{figure_example-1}. Considering the $Printing$ operation, it is  almost equivalent to $Printer$ and almost equivalent to $Impression$. We calculate the non-functional QoS degree of equivalence to determine which of $Printer$ or $Impression$ replace the best $Printing$.

First we calculate the values that we need for our degree computing. We detail the computing for $nbpage$.
\\
\\
$\begin{array}{llll}
& \mu(nbpage)\ =\ 56.66 & & \\
& \sigma(nbpage)\ =\ \sqrt{\left((60-56.66)^{2}\ +\ (100-56.66)^{2}\ +\ (10-56.66)^{2}\right)\ \div\ 3}\ =\ 36.84 & & \\
& z$-$score(nbpage_{printing})\ =\ (60-56.66)\ \div\ 36.84\ =\ 0.09 & & \\
& z$-$score(nbpage_{impression})\ =\ (100-56.66)\ \div\ 36.84\ =\ 1.176 & & \\
& z$-$score(nbpage_{printer})\ =\ (10-56.66)\ \div\ 36.84\ =\ -1.26 & & \\
& \eta(nbpage_{printing})\ =\ 0.477 && \\
& \eta(nbpage_{impression})\ =\ 0.206 & & \\
& \eta(nbpage_{printer})\ =\ 0.816 & & \\\\
& \eta(price_{printing})\ =\ 0.515 & & \\
& \eta(price_{impression})\ =\ 0.867 & & \\
& \eta(price_{printer})\ =\ 0.186 & & \\\\
& D_{sem}(access_{printing}, access_{printing})\ =\ 0,\ M_{Cpt}('wifi','wifi')\ =\ Exact & & \\
& D_{sem}(access_{printing}, access_{impression})\ =\ 0.2,\ M_{Cpt}('wireless','wifi')\ =\ PlugIn & & \\
& D_{sem}(access_{printing}, access_{printer})\ =\ 1,\ M_{Cpt}('bluetooth','wifi')\ =\ Fail & &
\end{array}$
\\
\\
\\
The $QoS_{Degree}$ of the three operations are:
\\
\\
$ QoS_{Degree}(Printing, Impression) = w1*(|\eta(nbpage_{printing})\ -\ \eta(nbpage_{impression})|)\ + w2*(|\eta(price_{printing})\ -\ \eta(price_{impression})|)\ + w3*(D_{sem}(access_{printing}, access_{impression}))$
\\
\\
$ QoS_{Degree}(Printing, Impression) = w1*0.27\ + w2*0.35\ + w3*0.2$
\\
\\
$QoS_{Degree}(Printing, Printer) = w1*(|\eta(nbpage_{printing})\ -\ \eta(nbpage_{printer})|)\ + w2*(|\eta(price_{printing})\ -\ \eta(price_{printer})|)\ + w3*(D_{sem}(access_{printing}, access_{printer}))$
\\
\\
$QoS_{Degree}(Printing, Printer) = w1*0.33\ + w2*0.33\ + w3*1$
\\
\\
If we suppose the three non-functional QoS properties of the same importance $w1 + w2 + w3 = 1$, we obtain:
$QoS_{Degree}(Printing, Impression) = 0.27$, and $QoS_{Degree}(Printing, Printer) = 0.55$. 
The $Impression$ operation offers non-functional QoS that are closer to $Printing$ than $Printer$ if we assign the same weight to the three non-functional properties.
\end{exple}

\section{Semantic Service Substitution in Pervasive Environments}

An application executing a service in pervasive environments would like to benefit from all the available services. Service substitution based on semantic interface matching and non-functional QoS properties is something the pervasive environment can provide to applications. We use the equivalence and almost equivalence relations to compare services together to know if one service can substitute another one. And we use the QoS degree equivalence to be sure that the services we provide to applications fit their needs. When a service appears in the environment, It can be functionally equivalent to another service being executed by an applications and with better QoS parameters. The environment will spontaneously substitute the service of the application with this new service. On the other hand, when a service disappear, the environment will look for equivalent or almost equivalent services with QoS properties similar to the vanishing services and redirect the application calls to this new service.
These two actions of spontaneously substituting services to applications allow these latter to execute properly despite the environment dynamicity.
\\
\\
\textbf{Service appearance}
\\
\\
Considering a set $S$ of finite services in the environment, we denote $si$ the service that appears. As a first step, the pervasive environment searches for functionally equivalent or almost equivalent services interfaces in the environment. Indeed, these services are services that provide the same functionality - the same functional interfaces - as the service $si$, and can be replaced in the application clients execution by the service $si$ . 

We consider the new service $si$. We suppose that the service $si$ is equivalent or almost equivalent to other services in the environment:
\\
\\
$\exists\ sj\ \in\ S$, $(\equiv_{sem}(si, sj)\ =\ true) \vee\ (\triangleright_{sem}(si, sj)\ =\ true)$
\\
\\
The spontaneous service $si$ substitution succeeds if $si$ can replace $sj$ for the application execution and that by providing better non-functional QoS properties than sj for the applications. By checking the profile of applications, the pervasive environment knows the values and the priorities ($wi$) that the applications would like to assign to the non-functional QoS properties. The environment can simulate a service $sk$, with these values, and calculates the $QoS_{degree}$ using the $wi$ specified by the applications. If no $wi$ are assigned, the pervasive environment applies the following values: $\sum_{i\ \in\ \N}wi\ =\ 1$. The service substitution succeeds if:
\\
\\
$QoS_{degree}(si,sk) < QoS_{degree}(sj,sk)$
\\
\\
which means that the new service $si$ is closer to $sk$ than $sj$ is to $sk$ in terms of non-functional QoS properties, $sk$ reflecting the applications needs and preferences for the non-functional QoS properties of the service they execute.

\begin{exple}
Considering the three operations defined in figure~\ref{figure_example-1}.

The $Printing$ service is a new service appearing in the environment and is semantic almost equivalent to the $Impression$ service. The environment considers applications using the $Impression$ service, and verifies which non-functional QoS properties are the required by the applications. For example, if the $price$ is important, the $w_{price}$ would be much more important than the $w_{access}$ and $w_{nbPage}$, and the new $Printing$ service fits better for the application.
The environment simulates a new service by assigning it the adequate values of the non-functional QoS properties required by applications. As an example we can give the following application required non-functional QoS properties depicted under service $sk$:

$\begin{array}{llll}
& Np_{QL} = \left\{<access,\ ``wireless''>\right\} & &\\
& Np_{QN} = \left\{<nbPage,\ 50,\ '>'\ >,\ <price,\ 12,\ '<'\ >\right\} & &
\end{array}$
\\
\\
And $w_{price}\ = 0.6,\ w_{access}\ =\ 0.2,\ w_{nbPage}\ =\ 0.2$
\\
\\
First we calculate the values that we need for our degree calculations:
\\
\\
The mean for $nbpage$ property: $\mu(nbpage)\ =\ 55 $
\\
The standard deviation for $nbpage$ property: $\sigma(nbpage)\ =\ 32$
\\
The normalised z-score values are: $\eta(nbpage_{printing})\ =\ 0.46$
\\
$\eta(nbpage_{impression})\ =\ 0.149$
\\
$\eta(nbpage_{printer})\ =\ 0.85$
\\
$\eta(nbpage_{sk})\ =\ 0.54$
\\
\\
The mean for $price$ property: $\mu(price)\ =\ 11 $
\\
The standard deviation for $price$ property: $\sigma(price)\ =\ 6,4$
\\
The normalised z-score values are: $\eta(price_{printing})\ =\ 0.46$
\\
$\eta(price_{impression})\ =\ 0.85$
\\
$\eta(price_{printer})\ =\ 0.15 $
\\
$\eta(price_{sk})\ =\ 0.539$
\\
\\
The semantic distance for the non-functional properties are:
\\
$D_{sem}(access_{printing}, access_{sk})\ =\ 0.8,\ M_{Cpt}('wifi','wireless')\ =\ Subsume$
\\
$D_{sem}(access_{impression}, access_{sk})\ =\ 0,\ M_{Cpt}('wireless','wireless')\ =\ PlugIn$
\\
$D_{sem}(access_{printer}, access_{printer})\ =\ 1,\ M_{Cpt}('bluetooth','wireless')\ =\ Fail$
\\
\\
Using these values we calculate:
\\
\\
$\begin{array}{llll}
& QoS_{degree}(Printing, sk)\ = 0.6*0.08\ +\ 0.2*0.8\ +\ 0.2*0.078\ =\ 0.22 & &\\
& QoS_{degree}(Impression, sk)\ = 0.6*0.391\ +\ 0.2*0\ +\ 0.2*0.311\ =\ 0.29& &
\end{array}$
\\
\\
We have $QoS_{degree}(Printing, sk)\ <\ QoS_{degree}(Impression, sk)$, which means that the new $printing$ service fits better the application requirements.
\end{exple}

\noindent
\textbf{Service disappearance}
\\
\\
Another major issue requiring service substitution is the disappearance of services form the environment. If a service disappear, the service registry of the environment is notified. This one asks the environment to come back with all the services that are equivalent or almost equivalent to this service. If many services are found, the environment creates sets of services. A set for the services equivalent and another one for the almost equivalence. The equivalence is considered better than the almost equivalence, as services can be interchanged in an equivalence relation (symmetric relation).

We denote $si$ the service that disappears and for this service the environment finds the equivalent or almost equivalent services:
\\
\\
$\exists\ sj\ \in\ S$, $(\equiv_{sem}(sj, si)\ =\ true) \vee\ (\triangleright_{sem}(sj, si)\ =\ true)$
\\
\\
We define the following:
\\
\\
$\begin{array}{llll}
& S_{\equiv}:\ set\ of\ sj,\ (\equiv_{sem}(sj, si)\ =\ true) & & \\
& S_{\triangleright}:\ set\ of\ sj,\ (\triangleright_{sem}(sj, si)\ =\ true) &&
\end{array}$
\\
\\
In every set, services are ordered following the $QoS_{degree}$ function that returns for every equivalent services with the service that disappeared their degree of equivalence concerning the non-functional QoS properties related to the service that the environment would like to replace.

By checking the values on the non-functional QoS properties for each service of every set, the environment calculates  the $QoS_{degree}(sj,si),\ \forall\ sj\ \in\ S_{*}$, of each service of a set with the service $si$. If no ponderation is given by the applications upon the priority of the properties the environment employs the same value for $wi$ : $\sum_{i\ \in\ \N}wi\ =\ 1$. 
The services within each set are ordered from the best one (service $sj$ that minimizes $QoS_{degree}(sj,si)$) to the worst one (service $sk$ that maximize $QoS_{degree}(sj,si)$):
\\
\\
$\begin{array}{llll}
& T_{\equiv}:\ set\ of\ ordered\ sj,(QoS_{degree}(s_{j},si)<QoS_{degree}(s_{j+1},si),j \in\ [1..|S_{\equiv}|-1]) & & \\
& T_{\triangleright}:\ set\ of\ ordered\ sj,(QoS_{degree}(s_{j},si)<QoS_{degree}(s_{j+1},si),j \in\ [1..|S_{\triangleright}|-1]) &&
\end{array}$
\\
\\
When a service $si$ disappears, the environment chooses the best replacement for the service $si$ by beginning from the most suitable set with the most suitable non-functional QoS properties.

\begin{exple}
Returning to our example of the $Printing$, $Impression$, and $Printer$ services (cf. figure~\ref{figure_example-1}).

If we search to replace the $Printing$ service because of a sudden disappearance and need to choose between the $Impression$ or the $Printer$ services, the calculated $QoS_{degree}$ between these services are different depending on the values assigned to $wi$.
\\
\\
$QoS_{Degree}(Printing, Impression) = w1*(|\eta(nbpage_{printing})\ -\ \eta(nbpage_{impression})|)\ + w2*(|\eta(price_{printing})\ -\ \eta(price_{impression})|)\ + w3*(D_{sem}(access_{printing}, access_{impression}))$
\\
\\
$QoS_{Degree}(Printing, Impression) = w1*0.27\ + w2*0.35\ + w3*0.2$
\\
\\
$QoS_{Degree}(Printing, Printer) = w1*(|\eta(nbpage_{printing})\ -\ \eta(nbpage_{printer})|)\ + w2*(|\eta(price_{printing})\ -\ \eta(price_{printer})|)\ + w3*(D_{sem}(access_{printing}, access_{printer}))$
\\
\\
$QoS_{Degree}(Printing, Printer) = w1*0.33\ + w2*0.33\ + w3*1$
\\
\\
If the service $Printing$ is no longer available, the environment finds the services $Impression$  and $Printer$ as almost equivalent to $Printing$.
For their non-functional properties, it is clear that if the environment assigns the same value to the three $wi$, the $Impression$ service would have a closer degree to $Printing$. Nevertheless, if the application using $Printing$ gives more importance to the price of the printing service, the environment will assign to $w2$ a greater importance, and we can notice the $Printer$ service has a closer degree to $Printing$ than the $Impression$ service.
\end{exple}

It can occurs that no equivalent or almost equivalent services are found, in that case the search may be refined over a set of operations. If the users and applications of the services that disappeared used a particular operation or set of operations, the search may be specified over these operations using the equivalence and almost equivalence service relations defined upon particular operations ($\equiv^{Op}_{sem},\ \triangleright^{Op}_{sem}$).
\\
\\
The spontaneous service $si$ substitution over a predefined set of operations $Op$ succeeds if:
\\
\\
$\exists\ sj\ \in\ S$, $(\equiv^{Op}_{sem}(sj, si)\ =\ true) \vee\ (\triangleright^{Op}_{sem}(sj, si)\ =\ true)$

\begin{exple}

Considering the three services interfaces and their semantic descriptions in figure~\ref{figure_ex-3}:
\\
\\
We have $ (\triangleright^{\left\{op1_{ifc1},\ op2_{ifc1}\right\}}_{sem}(ifc1,ifc2)\ =\ true)$, which means that the services proposing the interface $ifc1$ with the operations $op1_{ifc1}$ and $op2_{ifc1}$ can replace the operations $op1_{ifc2}$ and $op3_{ifc2}$ of service $ifc2$.
\end{exple}

As for the service as a whole, the environment requires to create the sets of equivalent and almost equivalent services over the predefined set of operations. It also orders the services within these sets depending on the non-functional QoS properties of the concerned operations and not the non-functional QoS properties of all the service.

If no services are found, the environment may consider the services that are $Subsume$ matching with the service that disappeared. If this replacement can fail to provide the required functionality as a $Subsume$ matching between services does not guarantee that the new service can provide all what the other service provided, it can allows the environment to provide something to the applications even if not exactly what is required, while awaiting the appearance of the desired services. The environment proposes these services to the applications, specifying that the services they seek are no longer available.

In case of complete failure of finding an appropriate service, the service registry of the environment redirects all the calls to the functional interface of the disappearing service to a proxy. Once a service registers a functional interface responding to the applications needs, the calls of the proxy can be redirected to this new service.

\section{Evaluation of the Semantic Service Substitution}

We implemented, as a proof of concept, all the major functionalities of the service substitution under an OSGi service platform implementation, the Apache Felix. The service semantic matching is done using online reasoner OWL-S ontologies (\cite{10}) and the matching relations of Paolucci (\cite{3}). The non-functional QoS properties are for now defined in the service description and we do not yet consider the dynamic changes affecting these properties while service execution. For the evaluations we developed a use case composed of 100 OSGi services in a small environment deployed on three laptops (Dell Latitude D410, 1,73 GHz, and 0,99 Go of memory).

\begin{figure}[!h]
\centering
\includegraphics[width=0.75\textwidth]{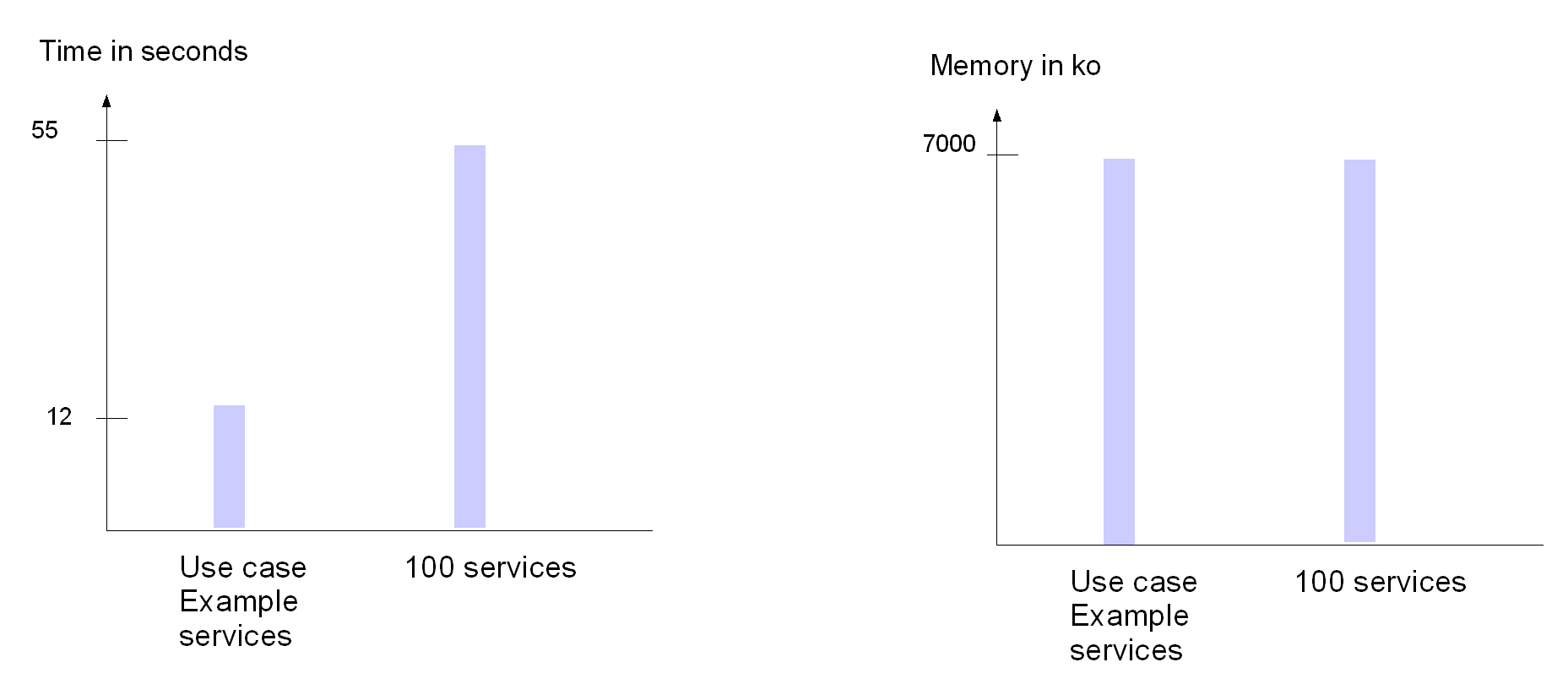}
\caption{Time execution for semantic service matching}
\label{figure_semantic}
\end{figure}

The semantic matching is  quite heavy (cf. figure~\ref{figure_semantic}). The OWL-S API takes about 12 seconds to compare and matches 8 services owl-s descriptions (\textsl{MyStudio}) and 55 seconds for about 100 services. The pellet matching engine that reads all the owl-s files by adding them to the reasoner and extracts the inputs, outputs and concepts fields is much slower and much more memory consumer than as simple syntactic matching based for example on introspection methods provided by the Java language. We conclude that the semantic matching using online semantic reasoning is a very heavy process.  We can improve the matching time and memory consuming by employing techniques as in PERSE (\cite{8}) that propose efficient semantic service matching using encoding classified ontologies.

\begin{figure}[!h]
\centering
\includegraphics[width=0.8\textwidth]{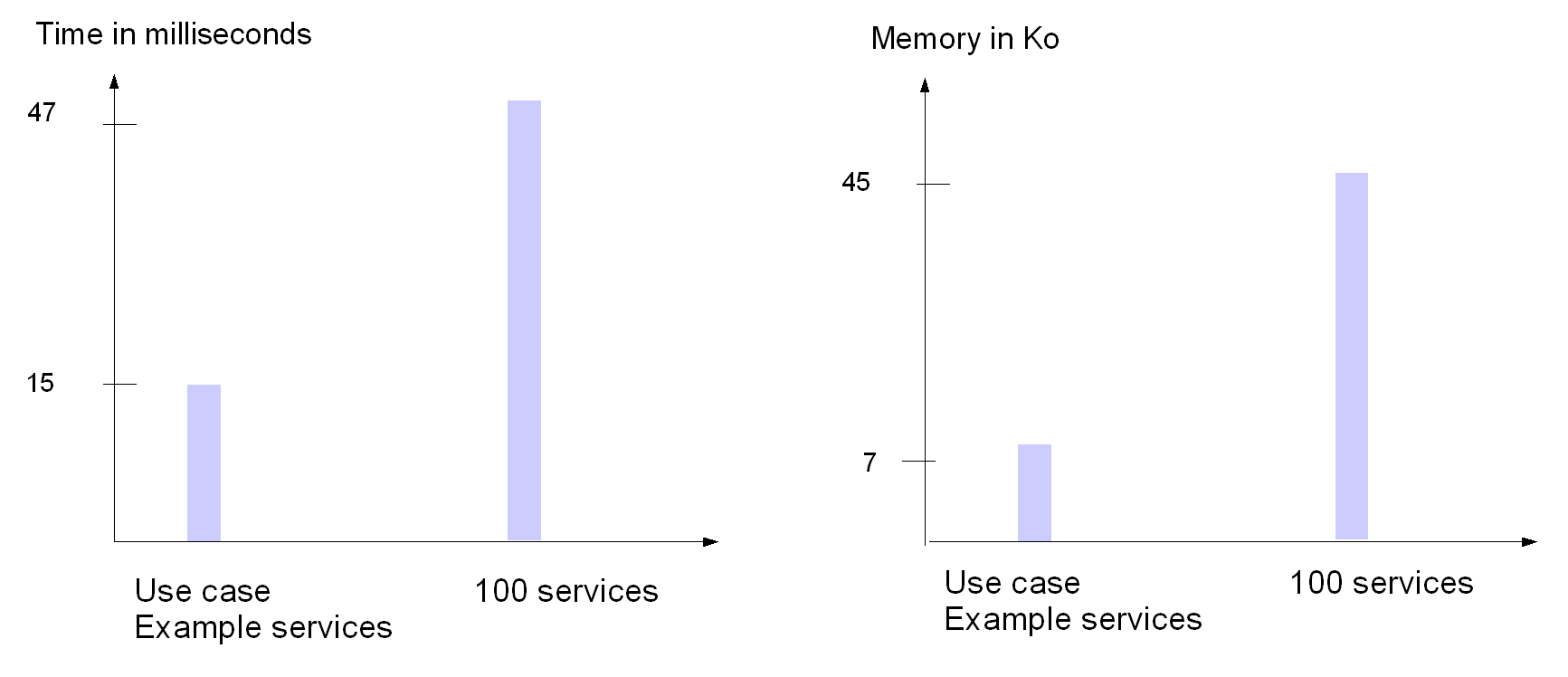}
\caption{Time and memory consumption for QoS degree computing}
\label{figure_QoS}
\end{figure}

Figure~\ref{figure_QoS} gives the time execution and memory consumption for quantitative non-functional properties $QoS_{degree}$ function computing. We suppose that each service has one quantitative non-functional property. When a service leaves the environment, the time to adapt to this changes is the time required to compute and sort the QoS degree of available services publishing the same interfaces (47 milliseconds for 100 services). When a service appears in the environment, the environment computes the QoS degree of this services to find if it better suits the applications using equivalent services. If so, the service registry will propose to applications the new service and the adaptation would be done in no time for the application, as it is showed figure~\ref{figure_QoS}. 

\section{Conclusion}

Service substitution is used in runtime reconfiguration in SOA systems in order to tolerate runtime variations and ensure continuity in service provisioning for the users.  Providing functionally equivalent services to the applications with better quality of services when services appear and disappear is a challenging problem as services are provided with different technologies and different characteristics. If many middleware proposed to semantically compare services and to adapt them to the application execution, few formalised and defined the service relations and especially the non-functional QoS properties degree metrics between services. We proposed a metric to compare services, based on semantic interface matching and a metric for computing the non-functional QoS property similarities between services.  We implemented a prototype under Java OSGi framework as a proof of concept and evaluated the efficiency of our proposal. 

One of the aspects that is not yet tackled by our middleware prototype is the state of a service (\cite{6}) that disappears while executing. If a service disappears while executing an application needs, to replace it in a transparent way, the environment needs not only to find equivalent services in terms of functional and non-functional QoS properties but to know from which state to start the execution of the new service, so that the application does not loose what has been already executed by the previous service. Mechanisms of logging and checkpoints need to be introduced at the service execution time level to save the state of a service at runtime. These mechanisms allow the environment to keep a trace over the state of services and to know when they disappear at which state of execution they were. Another important issue would be to test our prototype in large pervasive environments, such as university campus, were thousands of services may meet and where a real end user experience could be tested to evaluate the interest of our spontaneous service substitution approach vis à vis to users. Our approach would surely have problem to scale to these service numbers and a more smart selection, based not only on semantic ontologies but also on user profiles, would be appropriate to choose a subset of services to substitute.




\end{document}